\documentclass[prd,amsmath,amssymb,superscriptaddress,preprintnumbers,twocolumn,10pt,nofootinbib]{revtex4-1}

\usepackage{graphicx}
\usepackage{dcolumn}
\usepackage{bm}
\usepackage{amssymb}
\usepackage{latexsym}
\usepackage{booktabs}
\usepackage{amsmath}
\usepackage{multirow}
\usepackage{url}
\usepackage{float}
\usepackage[colorlinks=true, linkcolor=red, citecolor=blue]{hyperref}

\usepackage[normalem]{ulem}
\usepackage{color}
\usepackage{array}
\usepackage{enumerate}
\usepackage{subfigure}
\usepackage{hyperref}
\usepackage{graphicx}
\newcommand{\orcid}[1]{%
  \href{https://orcid.org/#1}{%
    \includegraphics[height=2ex, keepaspectratio]{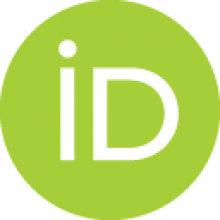}%
  }%
}

\begin{document}

\title{Assessing the robustness of amortized simulation-based inference to transient noise in gravitational-wave ringdowns}

\author{Song-Tao Liu~\orcid{0000-0001-5966-2461}}
\affiliation{Liaoning Key Laboratory of Cosmology and Astrophysics, \ College of Sciences, Northeastern University, Shenyang 110819, China}
\author{Tian-Yang Sun~\orcid{0009-0002-5109-6420}}
\affiliation{Liaoning Key Laboratory of Cosmology and Astrophysics, \ College of Sciences, Northeastern University, Shenyang 110819, China}
\author{Yu-Xin Wang~\orcid{0000-0002-9592-9087}}
\affiliation{Liaoning Key Laboratory of Cosmology and Astrophysics, \ College of Sciences, Northeastern University, Shenyang 110819, China}
\author{Yong-Xin Zhang~\orcid{0009-0001-5942-2991}}
\affiliation{Liaoning Key Laboratory of Cosmology and Astrophysics, \ College of Sciences, Northeastern University, Shenyang 110819, China}
\author{Shang-Jie Jin~\orcid{0000-0003-3697-3501}}
\affiliation{Liaoning Key Laboratory of Cosmology and Astrophysics, \ College of Sciences, Northeastern University, Shenyang 110819, China}
\author{Jing-Fei Zhang~\orcid{0000-0002-3512-2804}}
\affiliation{Liaoning Key Laboratory of Cosmology and Astrophysics, 
\ College of Sciences, Northeastern University, Shenyang 110819, China}
\author{Xin Zhang~\orcid{0000-0002-6029-1933}}\thanks{Corresponding author.\\zhangxin@mail.neu.edu.cn}
\affiliation{Liaoning Key Laboratory of Cosmology and Astrophysics, \ College of Sciences, Northeastern University, Shenyang 110819, China}

\affiliation{MOE Key Laboratory of Data Analytics and Optimization for Smart Industry, \ Northeastern University, Shenyang 110819, China}

\affiliation{National Frontiers Science Center for Industrial Intelligence and Systems Optimization, \ Northeastern University, Shenyang 110819, China}


\begin{abstract}
Gravitational waves (GW) emitted by binary systems allow us to perform precision tests of general relativity in the strong field regime. Ringdown signals allow for probing black hole mass and spin with high precision in GW astronomy. With improvements in current and next-generation GW detectors, developing likelihood-free parameter inference methods is crucial. This is especially important when facing challenges such as non-standard noise, partial data, or incomplete signal models that prevent the use of analytical likelihood functions. In this work, we propose an amortized simulation-based inference strategy to estimate ringdown parameters directly. Specifically, our method is based on amortized neural posterior estimation, which trains a neural density estimator of the posterior for all data segments within the prior range. The results show that our trained amortized network achieves valid confidence coverage and yields posteriors in agreement with established Markov-chain methods at moderate signal-to-noise ratios. At higher signal-to-noise ratios $(\rho \approx50)$, the recovered posteriors agree with the Markov-chain results in location, although the amplitude posterior is broader. The network provides inference speeds that are orders of magnitude faster.
Furthermore, we evaluate the robustness of the method against transient noise contamination. Our results indicate that the timing of glitch injection significantly affects estimation bias, in particular when the glitch contaminates the early, high-amplitude portion of the signal. Glitch strength is positively correlated with estimation error, but has limited effect at low glitch signal-to-noise ratios. The sensitivity is parameter dependent and varies with the glitch configuration. This study not only provides an efficient and statistically validated inference framework under the tested conditions for ringdown analysis but also lays a foundation for developing robust data-processing pipelines for future GW astronomy in realistic noise environments.

\end{abstract}

\maketitle
\section{Introduction}
    Gravitational waves (GW) were directly detected for the first time in 2015. The signal GW150914, which came from binary black holes merging, was discovered by LIGO~\citep{Abbott_2016}. Since then, GW astronomy has grown quickly. A growing number of GW events have been observed, including binary black hole mergers, binary neutron star mergers, and black hole-neutron star mergers~\citep{Abbott_2019,Abbott_2021_GWTC2,theligoscientificcollaboration2022gwtc21deepextendedcatalog,Abbott_2023,theligoscientificcollaboration2025gwtc40updatinggravitationalwavetransient}. A global detector network including LIGO, Virgo, and KAGRA has observed dozens of GW events. These observations strongly support Einstein's theory of general relativity. These observations also provide valuable insights into astrophysics, fundamental physics, and cosmology~\citep{Teukolsky:1973ha,Schutz1986,Holz_2005,Sathyaprakash_2009,Acernese_2014,2015,Abbott_2018,Annala_2018,Wang_2018,PhysRevLett.121.060506,Abbott_2019_170817}.  The ringdown phase of GW constitutes a crucial component of the GW signal emitted following the merger of compact objects, such as black holes or neutron stars~\citep{Abbott_2016}. Physically, this stage corresponds to the transition of the newly formed black hole toward a stable Kerr black hole state~\citep{Teukolsky:1973ha}. During this process, the remaining black holes are strongly perturbed and radiate GW to dissipate its initial energy and angular momentum, ultimately reach a stationary configuration~\citep{Berti_2009}. 
    
    Within the cosmological framework, general relativity serves as the cornerstone theory describing the evolution of large-scale structure, while the GW ringdown stage provides an experimental platform for testing this theory under extreme conditions~\citep{Berti_2015}. The characteristic signals of quasi-normal modes (QNM)~\citep{PhysRevD.76.064034} emitted during the ringdown phase are entirely determined by the mass and spin parameters of the newly formed black hole, allowing their frequency and damping rate to be used as precision tests of general relativity~\citep{Berti_2009,macarena_2024}.
    The ringdown signal can also serve as a standard measurement tool, providing an independent avenue for constraining cosmological parameters~\citep{Schutz1986,Cutler_1994,Guo_2019,Abbott_2023_Constraints,engelhardt2023detectingaxiondarkmatter,Song:2025ddm,Jin:2025dvf}. The measured QNM frequencies can calibrate the mass and spin of the final black hole, while these oscillation modes are closely linked to redshift effects~\citep{niccol_2022}. Consequently, observing the ringdown signals from distant black hole merger events can place constraints on cosmological parameters such as the Hubble constant, the dark energy equation of state, and so on in  broader cosmological models~\citep{emanuele_2006,Zhang:2012uu,Zhang:2015rha,Guo:2015gpa,abbott2017,Barack_2018,Guo:2018ans,Soares_Santos_2019,Hotokezaka2019,Zhang_2019,Fishbach_2019,ZHANG2019135064,Abbott_2021_Hubble,Li:2024qus,theligoscientificcollaboration2025gwtc40constraintscosmicexpansion,Song:2025bio}. 
    For example, the temporal evolution of ringdown data, combined with detectors of extremely high relative precision, such as Laser Interferometer Space Antenna (LISA)~\citep{amaroseoane2017laserinterferometerspaceantenna,Barack_2018} or third-generation GW observatories, offers a potential method for improving cosmic distance measurement~\citep{Jin:2025dvf, zhang2026factorizedneuralposteriorestimation}. Moreover, this approach allows for the construction of a cosmological black-hole population model through ``event by event'' detection, which can be used to study structure formation processes in the early universe~\citep{Zhang:2007bi,Zhu:2023gmx}.
    Furthermore, certain cosmological theories, such as dark energy models~\citep{Li:2012spm,Feng:2016djj,Li:2024qus} or scalar-tensor gravity~\citep{Gat_2013,Berti_2015,IOVINO2026140039}, may affect the speed and modal characteristics of GW propagation. By precisely measuring the QNM frequencies of the ringdown signal, it becomes possible to constrain the parameter space of such theories or test their validity.
    
    However, GW astronomy is still in the early stages. The available data are limited, and detector sensitivity remains insufficient~\citep{Abbott2020}. Progress depends on better theoretical models, and numerical simulations are a key tool in developing them~\citep{Pretorius_2005,GEORGE201864,Boyle_2019,KRASTEV2021136161,QIU2023137850}. Currently, traditional likelihood-based methods exhibit several limitations in the face of non-standard noise, partial data, and incomplete signal modeling~\citep{peiris2014considerationsinterpretationcosmologicalanomalies,Cesare_2020,xu2021optimizationgraphneuralnetworks}. Firstly, traditional likelihood methods typically assume that the noise is Gaussian and stationary, which simplifies the computation of the likelihood function~\citep{Fan_2019,Chua_2020,Gabbard_2021,isi2021analyzingblackholeringdowns}. However, actual GW detector data often contain non-Gaussian and non-stationary noise components, such as glitches~\citep{WEI2021136029,XU2024139016,gair2025contaminationtransientgravitationalwaves,Xiong_2025,theligoscientificcollaboration2026constraintsgravitationalwaves2024}. When noise characteristics deviate from these ideal assumptions, likelihood functions based on a Gaussian noise model may fail to accurately describe the data, leading to biased or inaccurate parameter estimates~\citep{6c61-fm1n}. Secondly, in some observational scenarios, only partial data may be available, or the data may contain gaps~\citep{Abbott_2020,Wurster_2022}. Traditional likelihood-based approaches can struggle to handle incomplete datasets, often requiring complex processing or interpolation techniques that may introduce additional errors or uncertainties~\citep{PhysRevD.105.102003}. Finally, accurate modeling of GW signals, particularly for strong field events such as black hole mergers, remains a challenging task~\citep{Pretorius_2005}. If the signal model is incomplete or inaccurate, for example by omitting significant QNM, even an accurate noise model may fail to infer the physical parameters correctly~\citep{PhysRevLett.129.111102}. 
    
    Due to the low event rate detected by current GW detectors, existing sequential neural network approaches have so far been limited to testing only the signal of an event~\citep{PhysRevLett.120.141103,Wei_2020,Cuoco_2021,Pacilio_2024}.
    However, with the advancement of GW detection technology, the next generation of GW detectors, including proposed detectors such as LISA, the Einstein Telescope (ET)~\citep{Punturo_2010}, and Cosmic Explorer (CE)~\citep{Abbott_2017,d._2022} will represent a revolutionary advance. The most fundamental and transformative improvement is the exponential increase in event rates~\citep{mauro_2022}. Compared to the current joint observations by LIGO/Virgo/KAGRA, which detect dozens of compact binary coalescence events per year, next-generation facilities will increase the observable event rate to thousands or even tens of thousands annually. Therefore, it is crucial to find an efficient parameter estimation method to deal with a large number of events~\citep{Cutler_1994,PhysRevD.97.044039,Dax_2021,Dax_2025}.
    
    Deep learning has garnered significant attention due to its successful application in GW detection~\citep{PhysRevD.97.044039,Chua_2020,Xia_2021,Jadhav_2021,PhysRevD.104.083021,Cuoco2022,PhysRevD.106.122002,Ma_2022,Sch_fer_2022,QIU2023137850,Nousi_2023,Zhao_2023,yun2023detectionextractionparameterestimation,PhysRevD.107.063029,Trovato_2024,stergioulas2024machinelearningapplicationsgravitational,McLeod_2025,zhong2026cosmologygalaxyclustersusing}. To overcome the computational limitations of traditional methods, researchers have begun exploring alternative approaches using deep learning to address these challenges. Neural Posterior Estimation (NPE) has emerged as a transformative approach within Simulation-Based Inference (SBI)~\citep{doi:10.1073/pnas.1912789117,hermans2020likelihoodfreemcmcamortizedapproximate,PhysRevLett.127.241103,Wang:2024oei}, offering a powerful alternative to traditional Bayesian methods for parameter estimation in complex scientific domains. Its development and application to GW astronomy~\citep{KRASTEV2020135330,maximilian_2021,uddipta_2023,thibeau_2024,tirapongprasert2026learningedgetaileduniformsampling}, particularly for ringdown analysis~\citep{Crisostomi_2023,Pacilio_2024}, represent a significant advancement in computational astrophysics~\citep{green2020completeparameterinferencegw150914}. NPE employing a deep learning-based variational autoencoder~\citep{s._2020,kingma2022autoencodingvariationalbayes,Sun:2025ypd}, takes advantage of the universal approximation capabilities of neural networks to directly learn a mapping from observed data to an approximate posterior distribution over model parameters, bypassing the need for an explicit, tractable likelihood function~\citep{papamakarios2018fastepsilonfreeinferencesimulation,Qin_2026}. This is especially advantageous in scenarios where the likelihood is intractable due to complex noise properties, incomplete signal models, or high-dimensional parameter spaces~\citep{Crisostomi_2023}. 
    The foundational concept of NPE is rooted in the broader field of SBI, also known as likelihood-free inference. In this paradigm, a simulator, a forward model that can generate synthetic data given a set of parameters, is used to create a large training dataset of input parameters and corresponding simulated observations. A neural network, often a conditional density estimator like a Mixture Density Network or a Normalizing Flow model, is then trained on this dataset to invert the simulator, effectively learning the posterior distribution. Once trained, this neural network can perform inference on new, real observational data in a fraction of the time required by traditional sampling-based methods like Markov Chain Monte Carlo (MCMC)~\citep{f._2009}. This speed-up is crucial for the era of big data in astronomy, where facilities like the Vera C. Rubin Observatory and next-generation gravitational wave detectors are expected to produce vast volumes of data. 
    
    Moreover, in the presence of non-Gaussian noise, inference using simple neural networks suffers from inherent limitations. Due to their limited robustness, such models may introduce significant bias when approximating the posterior distribution, particularly in the estimation of marginal parameter distributions and joint credible regions~\citep{hermans2020likelihoodfreemcmcamortizedapproximate,alex_2024}. To address this challenge, we developed and evaluated a deep learning framework based on amortized NPE for the rapid and accurate parameter estimation from GW ringdown signals contaminated by non-Gaussian noise. We propose an enhanced amortized NPE architecture featuring increased model complexity to improve the robustness of amortized inference~\citep{Dax_2025}. Specifically, our network is built upon a multi-layer perceptron augmented with residual connections~\citep{kaiming_2015}, forming an embedding-enhanced NPE estimator. The input data are first compressed into a low-dimensional latent representation via a residual connections embedding network. This compact embedding is then fed into highly flexible Normalizing Flows, specifically, a Neural Spline Flow (NSF), to enable expressive posterior approximation. This design significantly enhances the overall robustness of the inference pipeline. Crucially, the proposed NPE model undergoes rigorous statistical validation. We assess the calibration of posterior estimates using Monte Carlo coverage tests~\citep{hermans2022trustcrisissimulationbasedinference} and further evaluate their statistical consistency by applying the Kolmogorov-Smirnov (KS) test~\citep{2006Pattern} on a test set. Additionally, we test the model's performance on data contaminated with realistic simulated glitches and quantify its robustness by computing the Jensen-Shannon Divergence (JSD) between posterior estimates from glitch-contaminated and noise-free signals.
    
    
    This paper is organized as follows. In Section~\ref{sec2} we introduce the amortized NPE method, the ringdown waveform model, the data generation and training process, and the statistical tests used for model evaluation. Section~\ref{sec3} presents the calibration performance of the model on clean data and systematically analyzes the impact of transient noise, i.e., glitches on inference results. Finally, Section~\ref{sec4} summarizes the main findings and outlines future research directions.
    
\section{Methodology}\label{sec2}

    \subsection{Amortized NPE }\label{sec2.1}

        Amortized NPE commonly employs Normalizing Flow~\citep{Tabak2010,Tabak2013} as density estimators~\citep{Sun:2023vlq,Sun:2024ywb}. Normalizing flow uses an invertible bijective transformation $f$ to map a complex target distribution to a simple base distribution $\varpi(\boldsymbol{z})$ that is fast to evaluate. For amortized NPE, the target distribution is $p(\boldsymbol{\theta}|\boldsymbol{x} )$ and $\varpi(\boldsymbol{z})$ which is typically a simple base distribution. The transformation $ f: \boldsymbol{z}\rightarrow \boldsymbol{\theta}$ must be invertible and have a tractable Jacobian.  This is so that we can evaluate the target distribution from $\varpi(\boldsymbol{z})$ by a change of variable:

        \begin{equation}
            p(\boldsymbol{\theta}|\boldsymbol{x})=\varpi(\boldsymbol{z})\left|\operatorname{det} \left(\frac{\partial f^{-1}}{\partial \boldsymbol{\theta}}\right)\right| .\tag{1}
        \end{equation}

        The goal of amortized NPE is to estimate the posterior density over the entire prior volume covered by the training set. This means it trains a single neural network model that can estimate the full posterior distribution for any observed data drawn from the same prior distribution, without the need for retraining or performing inference anew for each individual observation. 

        We use amortized NPE to estimate posterior densities. Amortized NPE was introduced in~\citep{radev2023bayesflowamortizedbayesianworkflows,andrew_2025} and further expanded in astrophysics~\citep{Hahn_2022,Crisostomi_2023,kolmus2024tuningneuralposteriorestimation}. The aim of amortized NPE is to train an approximate density estimator $q_\phi \left(\boldsymbol{\theta}|\boldsymbol{x}\right)$ of the posterior density $p\left(\boldsymbol{\theta}|\boldsymbol{x}\right)$. Here $\boldsymbol{x}$ is the data segment and $\boldsymbol{\theta}$ is the set of model parameters to be estimated; to be more precise, $\boldsymbol{x}$ is the sum of  a simulated ringdown waveform data $\boldsymbol{h}\left(\boldsymbol{\theta}\right)$ which is the deterministic component and a stochastic component $\boldsymbol{n}$ (the noise of detector), i.e. $\boldsymbol{x}\left(\boldsymbol{\theta}\right) = \boldsymbol{h}\left(\boldsymbol{\theta}\right) + \boldsymbol{n}$. In brief, we generate $\boldsymbol{h}\left(\boldsymbol{\theta}\right)$ through a numerical process, so as to prepare a training set of paired samples $\left\{\boldsymbol{\theta_i},\boldsymbol{x_i}\right\}$  to train the density estimator.

        The training process maximizes the likelihood $\prod_i q_{\phi}\left(\boldsymbol{\theta}|\boldsymbol{x}_i \right)$ with respect to the model parameters $\phi$. The estimator asymptotically converges to the true posterior density as the size of the training set increases. As mentioned above, our estimator uses NSF~\citep{Qin_2026}, spline-based neural normalizing flows. Once trained, the estimator $q_{\phi}$ can be sampled from within a few seconds, returning a set of posterior samples $\{\boldsymbol{\theta}_n\}$.

    \subsection{Ringdown waveform}\label{sec2.2}

        The form of the GW strain can be expressed by
        \begin{equation}
            \boldsymbol{h}\left(\theta\right) = F_+h_+ + F_{\times}h_{\times}.
            \tag{2}
        \end{equation}
        Here $F_+$ and $F_{\times}$ are the pattern functions of the detector~\citep{PhysRevD.58.063001}, depending on the sky position and relative orientation of the source, the polarization angle of the waveform and the starting time of the ringdown.
        
        The plus mode and cross mode waveform are expressed as superpositions of damped sinusoids as 
        \begin{equation}
            h_+ = \sum_{l,m,n}{\mathcal{A}_{lmn}e^{-\frac{(t-t_{start})}{\tau_{lmn}}}\cos\left(\Phi_{lmn}\right)Y^+_{lm}(\iota)}, \tag{3a}
        \end{equation}
        \begin{equation}
            h_{\times} = \sum_{l,m,n}{\mathcal{A}_{lmn}e^{-\frac{(t-t_{start})}{\tau_{lmn}}}\sin\left(\Phi_{lmn}\right)Y^{\times}_{lm}(\iota)}. \tag{3b}
        \end{equation}
        Note that this formulation assumes equatorial symmetry, valid for non-precessing binaries. Precessing systems would require additional cross terms.
        
        For parameter estimation, we fix the sky position, polarization, starting time, and inclination as known parameters.

    \subsection{Data generation and training}\label{sec2.3}

        We perform data generation with only detector noise to test the ability to recover a set of known injection parameters. We simulate the simplest harmonic superposition model: $Kerr_{2,2,0}$, a system containing only the (2, 2, 0) mode. 

        For convenience, we adopt the same right ascension $\alpha$, declination $\delta$, and polarization $\psi$ parameters as those of the GW150914-like event, and fix the inclination angle at $\iota=\pi$~\citep{PhysRevLett.123.111102,PhysRevLett.129.111102}. 
        The noise is randomly generated from the designed power spectral density (PSD) of aLIGO. 
        For the simulated injections, we sample the strains at 2048 Hz and truncate the waveform to a duration of 0.1 seconds, thus resulting in a data segment of 204 bins. All ringdown signals are simulated for a single detector (LIGO-Hanford).

        Additionally, we simulate glitches using a deep generative model. Specifically, we employ the pre-trained Consistency-Term Wasserstein Generative Adversarial Network (CT-GAN) model provided in the open-source {\tt gengli} package, which implements a CT-GAN trained on real LIGO noise data from the second observing run (O2)~\citep{Lopez_2022}. We use the pre-trained weights for the LIGO-Hanford (H1) detector and generate glitches by sampling from the network’s latent space, following the default inference procedure described in the package documentation. The generated glitches are whitened and evaluated on a time grid sampled at 4096 Hz, then resampled to 2048 Hz and scaled to the desired SNR to match our data generation pipeline.
        
        According to the training data specification in the reference, the generated glitches have a fixed duration of 0.01 seconds concentrated in the frequency band of [30, 250] Hz and are sampled at 2048 Hz. This approach allows us to simulate the complex, non-Gaussian morphology of real blips that are often difficult to capture with simple analytical functions.
        
        We model the density estimator as an NSF~\citep{papamakarios2021normalizingflowsprobabilisticmodeling} and use the implementation from the {\tt lampe} package. The relevant hyper-parameters are listed in Table~\ref{tab1}. In particular, the NSF is a flow of 5 transforms, each spline-coupling layer modulated by a conditional network with 4 hidden layers and 128 hidden features.
        
        The NSF does not directly feed whitened data into the network for training. Instead, an embedding network is applied for dimensionality reduction: the simulated single-detector input is first up-sampled to twice its original dimension. The data are then mapped by a fully connected neural network to 128 bins, further compressed to 64 bins, and finally the compressed data are passed to the NSF for inference. This network follows an ``expand then compress'' strategy, consisting of three hidden layers. Each layer is equipped with layer normalization and dropout regularization to enhance training stability. 
        
        Subsequently, this embedded representation is used as a conditional input to a posterior estimator based on NSF. Here, five stacked spline-coupling layers perform invertible transformations modulated by corresponding conditional networks. The final output is the posterior probability density of the target parameters.
        
        We train and evaluate the density estimator on a batch of 4096 samples. In each round of training we generate 2000000 training samples. We use the {\tt Adam} optimizer with a learning rate of $0.001$. We train the posterior density estimators on a single GPU RTX A6000. The training runs take 200 epochs and last $9.6$ hours. The training was terminated automatically based on the training loss, rather than running the full 200 epochs, if no improvement was observed within the patience period.
        
        \begin{table}[ht]
        \centering
        \setlength\tabcolsep{10pt}
        \renewcommand{\arraystretch}{1.2}
        \caption{The key hyperparameters of the amortized NPE.}
        \small
        \begin{tabular}{cc}
                \hline \hline 
                \multicolumn{2}{c}{NSF}\\
                \hline 
                hidden features& $128$ \\ 
                transforms& $5$\\
                Activation & ReLU \\
                Batch Norm & False \\ 
                \hline 
                
                \hline \hline
                \multicolumn{2}{c}{Embedding network}\\
                \hline 
                input dimension&$408$\\
                hidden layers&$3$\\
                hidden dimension&$[200, 100, 128]$\\
                output dimension&$64$\\
                \hline 
                
                \hline \hline
                \multicolumn{2}{c}{Training hyper-parameters}\\
                \hline
                number of simulations&$2000000$\\
                batch size&$4096$\\
                learning rate&$0.001$\\
                patience&$50$\\
                varying noise&TRUE\\
                \hline \hline
            \end{tabular}
    
        \label{tab1}
        \end{table}

        In this study, we use the best signal-to-noise ratio (SNR)~\citep{Cutler_1994}. Although the definition in Ref.~\citep{Cutler_1994} is formulated in the frequency domain, it would be equivalent to the time-domain inner product via Parseval's theorem in the idealized case of a smoothly tapered, finite-duration signal. In practice, we compute this quantity by taking the fast Fourier transform of the time-domain strain $h(t)$ and evaluating the noise-weighted integral\footnote{The ringdown waveform starts abruptly and is not tapered at the lower-frequency end, so a direct FFT can produce spectral leakage~\citep{isi2021analyzingblackholeringdowns}. Therefore, the frequency-domain SNR is not strictly equivalent to the time-domain inner product via Parseval’s theorem. The SNR is used here only as a bookkeeping parameter for injection loudness, and our conclusions do not rely on this strict equivalence.}
        \begin{equation}
            \rho = \sqrt{4 \int_{0}^{\infty}{\frac{\left | \tilde{h}\left(f\right) \right |^2}{S_{n}\left(f\right)}df}}=\sqrt{\left \langle h  | h  \right \rangle } , \tag{4}
        \end{equation}
        where $\tilde{h}(f)$ is the Fourier transform of the ringdown strain $h(t)$, and $S_n(f)$ is the one-sided power spectral density of the detector noise. For the test injections, the fundamental-mode amplitude is set to $\mathcal{A}_{220}$, which yields a target SNR of approximately 12 for the chosen detector PSD.

        We evaluate the amortized NPE method using test data of simulated ringdown signals. For this test, the mass and spin of the remnant black hole are fixed to $\mathcal{M}_f = 67 M_{\bigodot}$ and $\chi_f = 0.67$. 
        We simulate realistic detection conditions comparable to observed events by performing test injections with a target SNR of approximately 12 for the glitch-free ringdown signal. This target SNR is achieved by setting the amplitude of the fundamental mode to $\mathcal{A}_{220}=2\times 10^{-21}$ within the corresponding noise PSD of the detector. All parameters are listed in Table~\ref{tab2}.

        \begin{table}[ht]
        \centering
        \setlength\tabcolsep{12pt}
        \renewcommand{\arraystretch}{1.2}
        \caption{Parameters of the simulated injections in test data.}
        \small
        \begin{tabular}{cc}
            \hline \hline 
            Parameter & Values \\
            \hline 
            Total mass& $\mathcal{M}_f = 67 M_{\bigodot}$\\
            Spin& $\chi_f = 0.67$\\
            Amplitude& $\mathcal{A}_{220}=2\times 10^{-21}$\\
            Phase& $\phi_{220}=\pi$\\
            \hline \hline
        \end{tabular}
        
        \label{tab2}
        \end{table}
        
        We draw posterior inferences using box-uniform priors on the model parameters, with the prior ranges listed in Table~\ref{tab3}. We benchmark our results against posterior samples obtained with the time-domain inference software {\tt pyRing (v2.3.0)}~\citep{PhysRevLett.123.111102,Carullo_2019,Abbott_2021,pyRing}. We sample with the nested sampler {\tt cpnest}, using 4096 live points and 4094 maximum Markov-chain steps, which typically results in about 20000 posterior samples. For consistency, we also extract 20000 posterior samples from the trained estimator when doing the comparisons.
        
        \begin{table}[ht]
        \centering
        \setlength\tabcolsep{12pt}
        \renewcommand{\arraystretch}{1.2}
        \caption{Priors of simulated ringdown waveform parameters. Note that other parameters not mentioned are set to zero for simplicity.}
        \small
        \begin{tabular}{cc}
            \hline \hline 
            Parameter & Uniform distribution \\
            \hline 
            Total mass& $\mathcal{M}_{f}\in[20,300]~ M_{\odot}$\\
            Spin& $\chi_f\in [0,0.99]$\\
            Amplitude& $\mathcal{A}_{220}\in[0.1, 50]\times10^{-21}$\\
            Phase& $\phi_{220}\in[0,2\pi]$\\
            \hline \hline
        \end{tabular}
        
        \label{tab3}
        \end{table}

    \subsection{Analytical method}\label{sec2.4}
        The KS test~\citep{Lopes2011} is a statistical method used to assess the quality of fit between a sample distribution and a reference distribution. In GW ringdown parameter estimation, the KS test can evaluate the consistency between the posterior distribution of parameters and a theoretical or predictive distribution~\citep{Sidery:2013zua,Veitch:2014wba,Biwer:2018osg,Thrane_2019,yu2022surfacedefectdetectionevaluation}. The test first calculates the cumulative distribution functions (CDF) of the two distributions. It then computes the distance $D_{\rm {n}}$ between the two distributions based on the CDF, which can be written as 
        \begin{equation}
            D_{n} = \max \left| F_{n}(x)-F(x)\right|,\tag{5}
        \end{equation} 
        where $F_{n}(x)$ is the empirical cumulative distribution function of the sample and $F(x)$ is the reference (theoretical) cumulative distribution function. 
        The resulting $D_{n}$ is also known as the KS statistic. The p-value, calculated based on the KS statistic, is generally employed as a direct measure to evaluate the similarity or difference  between two distributions. The p-value is commonly compared to the significance level. To be specific, if the p-value is less than the significance level, it indicates that the two distributions are considered statistically different. Here, the significance level is set to a value of $0.05$.
        
        In our work,  we compare the distribution of measured coverage probabilities against the theoretically expected uniform distribution to assess the calibration of the amortized NPE model. It can be shown~\citep{talts2020validatingbayesianinferencealgorithms,Karchev_2022} that, if the distribution generated by our model approximates the true posterior, then the expected coverage is distributed uniformly within $[0, 1]$, which is a uniform distribution.

        To assess the statistical reliability of the amortized NPE, we use the expected-coverage diagnostic implemented in \texttt{lampe.diagnostics} module of \texttt{lampe}. For each of the $N=100$ test injections we evaluate the posterior density at the true parameter vector and compute the fraction of posterior samples drawn from the trained estimator with density below this value, i.e., the expected-coverage statistic of the joint four-dimensional posterior. The empirical expected coverage is then compared against the nominal credible levels in a p-p plot (Fig.~\ref{coverages_all}). For a well-calibrated model this coverage statistic is uniformly distributed in $[0,1]$, which we verify with a KS test on the coverage levels; the corresponding $p$-value is given in Fig.~\ref{coverages_all}.
        
        
        The JSD quantifies the difference between two probability distributions and measures the similarity of two distributions~\citep{rao1987differential,61115}. It is achieved by taking a weighted average of the probability density functions of the two distributions, given by
        \begin{equation}
            {\rm JSD}(p\parallel q )=\frac{1}{2} {\rm KLD}(p \parallel \frac{p+q}{2} )+\frac{1}{2} {\rm KLD}(q\parallel \frac{p+q}{2} ),\tag{6}
        \end{equation}
        where Kullback-Leibler divergence (KLD)~\citep{10.1214/aoms/1177729694,10.1214/aop/1176996454} is defined as
        \begin{equation}
           {\rm KLD}(p\parallel q)=\int_{-\infty }^{+\infty } p_{(x)}\ln_{}{} \frac{p_{(x)} }{q_{(x)} } \mathrm{d}x.\tag{7}
        \end{equation}
        
        In practice, we adopt the discrete form of KLD rather than the continuous integral form Eq. (7). This is because the posterior samples obtained via MCMC or density estimators are inherently discrete. Directly using the discrete summation is not only computationally efficient, but also naturally aligns with the sampling representation of our estimated distributions.
        \begin{equation}
            {\rm KLD}(p\parallel q)=\sum_{k=1 }^{K } p_{k}\ln_{}{} \frac{p_{k} }{q_{k} } .\tag{8}
        \end{equation}

        To evaluate the discrete sum in Eq. (8), we represent each posterior distribution as a discrete sample set and estimate the densities $p_k$ and $q_k$ from histograms on a common grid. For the marginal JSD, we construct shared one-dimensional histograms of the two sample sets with 20 equal-width bins covering the combined range of the samples. For the joint four-dimensional JSD, we use 10 equal-width bins per dimension. The normalized histogram counts define $p_k$ and $q_k$ at the corresponding grid points, and the JSD is then evaluated using Eqs. (6) and (8) with base-2 logarithms. This approach directly uses the sampling representation of the posteriors and avoids introducing additional density estimation assumptions. In the glitch studies, $p$ and $q$ denote the posteriors inferred by the amortized NPE from clean data and glitch-contaminated data, respectively.
        
        While the KLD can measure the difference  between two probability distributions, it is asymmetric because the KLD between distributions $p$ and $q$ is not equal to that between $q$ and $p$. To overcome this asymmetry,  we employ the JSD, which provides a symmetric and smoothed measure. The JSD is bounded between $0$ and $1$ for base-2 logarithms, where $0$ indicates identical distributions and 1 corresponds to maximal dissimilarity~\citep{Romero-Shaw:2020owr}. This bounded range facilitates the interpretation and comparison of distributional differences. In this work, we use the JSD to quantify the discrepancy between the parameter posterior inferred by the amortized NPE model from data contaminated with glitches and the ideal posterior.
      
\section{Results and discussion}\label{sec3}

    \subsection{Amortized NPE model evaluation }\label{sec3.1}

        Before assessing the model's robustness against glitches, we first evaluate the accuracy of the amortized NPE model using data free of non-stationary noise. In this section, we systematically evaluate the performance of the amortized NPE model for parameter inference of GW ringdown signals without glitches through numerical experiments. 
        To investigate how training epochs affect the prediction credibility of the amortized NPE, we compare models trained for a small number of epochs (100 epochs) against those trained for a large number (200 epochs), evaluating their posterior coverage across multiple credible intervals. 
        
        \begin{figure*}[htbp]
        	\centering
        	\subfigure[low-epoch]
        	{
        		\begin{minipage}[b]{0.48\linewidth}
        			\centering
        			\includegraphics[scale=0.65]{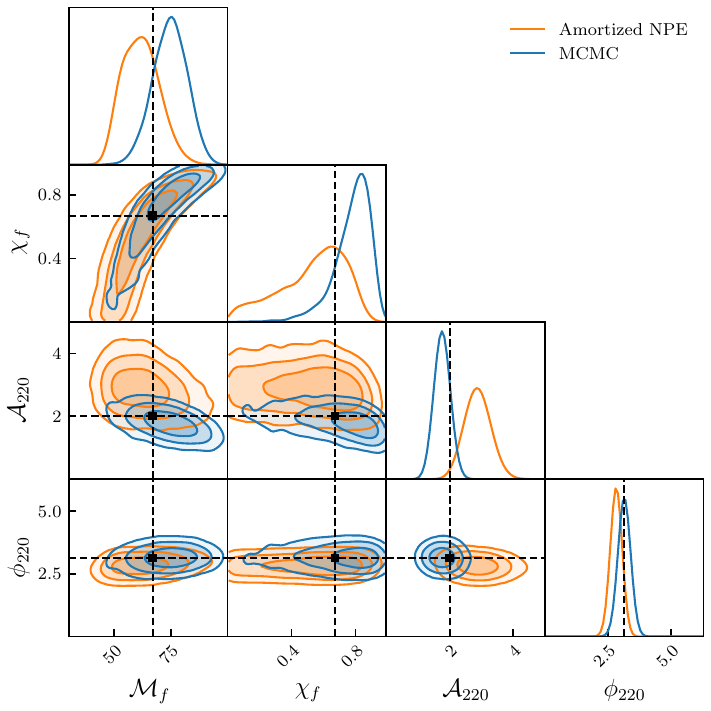}
        		\end{minipage}
        	}
        	\subfigure[high-epoch]
        	{
        		\begin{minipage}[b]{0.48\linewidth}
        			\centering
        			\includegraphics[scale=0.65]{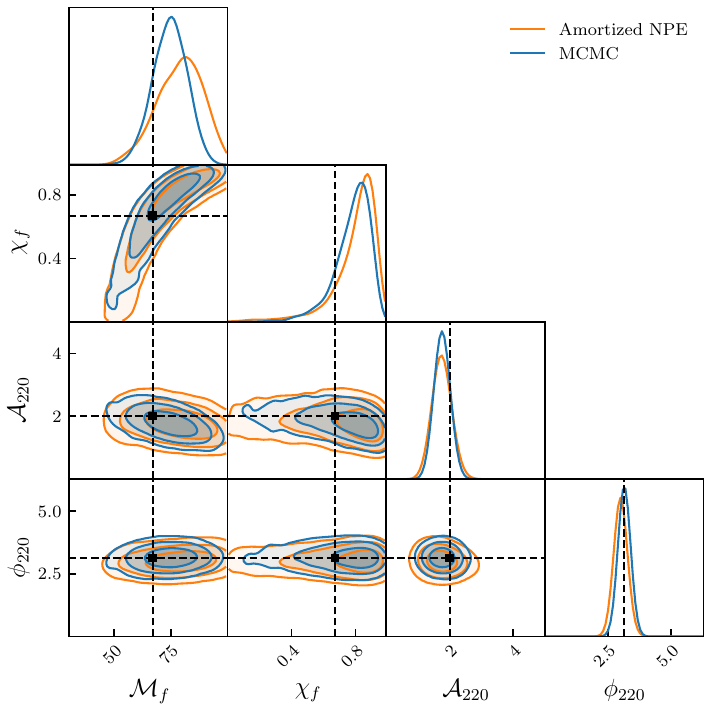}
        	\end{minipage}
        	}
        	\caption{Comparison of parameter estimates and confidence intervals derived from MCMC and amortized NPE. The figure shows the mean values for parameters $\mathcal{M}_{f}$, $\chi_f$, $\mathcal{A}_{220}$, and $\phi_{220}$ under the two methods, alongside their $68.3\%$, $95.5\%$, and $99.7\%$ credible intervals from the full posterior distribution. The black line marks the values of the injection parameters.}
        \label{corner_all}
        \end{figure*}
        
        With limited training, the model initially learns the basic structure of the posterior distribution.
        Compared to the results from traditional nested sampling, the central tendencies of the key parameter posterior distributions inferred by the model are largely consistent. However, slight deviations exist in the inferred confidence intervals. The left panel of Fig.~\ref{corner_all}  compares parameter estimates and confidence intervals from MCMC and amortized NPE, showing a slight mismatch in the inferred credible regions across different confidence levels. 
        
        After sufficient training, model performance improves significantly. The posterior distributions it infers show high visual agreement with MCMC results, and divergence measures quantified by the KS test drop to very low levels $(p \geq 0.05)$. The results are shown in the right panel of Fig.~\ref{corner_all}. This demonstrates that a well-trained amortized NPE model produces posterior distributions that are statistically consistent with high-precision MCMC results. At the same time, it achieves inference speed-ups of several orders of magnitude. 
        
\begin{figure}[htbp]
        \centering
        \includegraphics[width=0.38\textwidth]{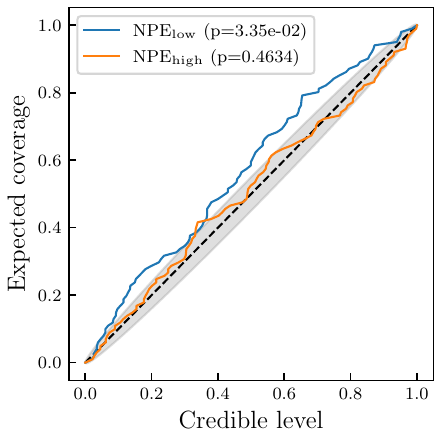}
        \caption{The nominal credible level and the expected coverage of the joint posterior computed with the \texttt{lampe} expected-coverage diagnostic. The blue curve represents the 100-epoch model, while the orange curve corresponds to the 200-epoch model. The grey shaded band indicates the $1\sigma$ counting uncertainty and a perfectly calibrated model would lie on the diagonal black line. }
        
        \label{coverages_all}
        \end{figure}

        As internal diagnostics, we further quantify the statistical reliability of the uncertainty estimates through coverage probability tests. The expected coverage distributions for both training processes are summarized in Fig.~\ref{coverages_all}. 
        To assess the statistical significance of the deviation from the ideal diagonal, we consider the expected counting uncertainty. The p-p plot is constructed using N=100 independent injections.
        
        For the low-epoch model, the empirical expected coverage significantly deviates from the ideal line, particularly in the tails. Consequently, the distribution fails the KS test. This indicates that the model is not fully calibrated, and its posterior uncertainties may be either underestimated or overestimated. In contrast, the high-epoch model shows nearly ideal calibration. Its expected coverage closely follows the diagonal reference line, and the corresponding p-values are statistically indistinguishable from a uniform distribution. This demonstrates that the well-trained posterior is consistent with satisfactory calibration over the tested parameter range.
        
        We additionally repeat the comparison at $\rho \approx 50$ ($\mathcal{A}_{220}=8.3\times10^{-21}$). The posterior distributions inferred by the amortized NPE agree in location with the MCMC results for the final mass $\mathcal{M}_{f}$, the spin $\chi_f$, and the phase $\phi_{220}$. For the amplitude $\mathcal{A}_{220}$, the NPE and MCMC posteriors agree in location, but the NPE posterior is somewhat broader. We therefore regard the high-SNR comparison as exercising the pipeline at an additional, higher-SNR configuration rather than as an exact-match test. This indicates that the amplitude parameter is the most sensitive parameter of the estimator at this SNR. The corner plots are presented in Appendix~\ref{app:snr50}.

    \subsection{Influence of glitches on model }\label{sec3.2}

        Non-Gaussian noise refers to a class of noise processes whose statistical distribution deviates from the Gaussian or normal probability model. 
        This noise exhibits more complex statistical characteristics and generative mechanisms, which can significantly compromise the stability of the system, degrade the accuracy of the inference, and reduce the analytical efficiency.
        In high-precision observational fields like GW astronomy and astrophysical data analysis, non-Gaussian noise, especially glitches, poses a major practical challenge. It directly compromises data interpretability and undermines the reliability of scientific conclusions. 
        
        \begin{figure*}[htbp]
        \centering
        \includegraphics[width=0.8\textwidth]{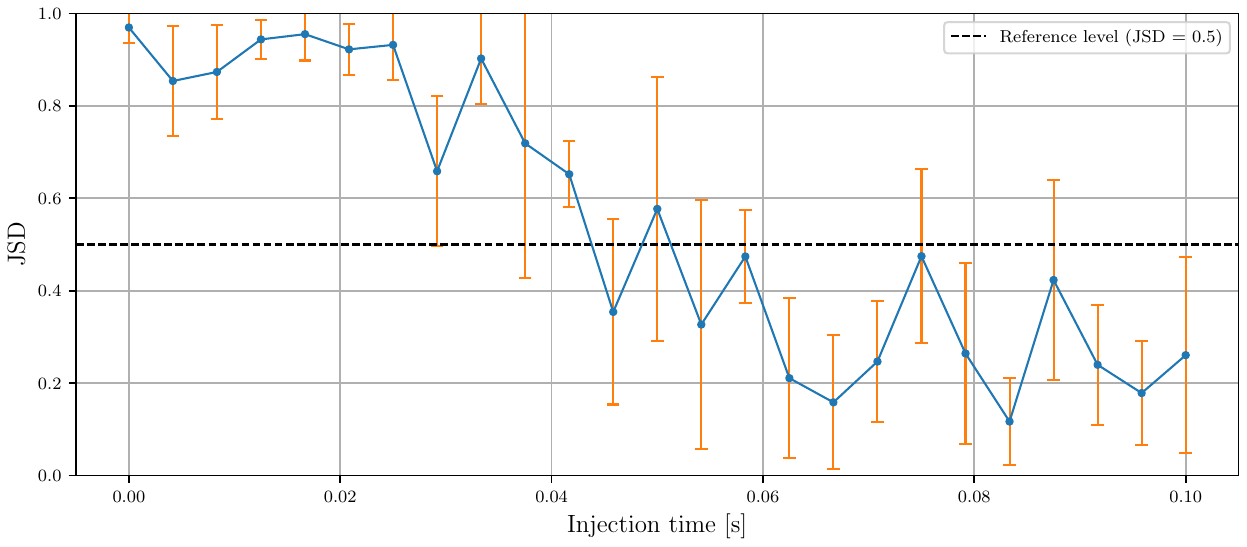}
        \caption{The JSD values between posterior distributions obtained from data contaminated with glitches and those from clean data at the time point of noise insertion. The data points in the graph are the mean JSD calculated from five independent runs. Error bars represent the standard deviation (SD) of these five runs. The horizontal dashed line at JSD = 0.5 is shown only as a visual reference and is not interpreted as a formal statistical-significance threshold.}
        \label{Time_ALL}
        \end{figure*}
        In this section, we evaluate the effect of non-Gaussian noise (glitches) on the posterior distribution inferred by the model at different time points and for different SNR of the signal. To assess the robustness of the inference pipeline, we inject simulated glitches into ringdown signals and analyze the resulting biases in the posterior distributions. The following presents our results of glitch injection and the key findings on model performance degradation. 
    
    \subsubsection{The time point of glitches }\label{sec3.2.1}
        In this section, we assess the impact of glitch injection time on model reliability by measuring the JSD value between the posteriors. 
        For convenience, we choose glitches with SNR $=$ $14$ for non-Gaussian noise injection.
        Figure~\ref{Time_ALL} quantifies how the timing of glitch injection affects the reliability of parameter inference. We measure this impact using the JSD, a symmetric and bounded metric for distributional discrepancy. Here, a JSD value of 0 indicates identical posterior distributions, while larger values correspond to greater deviation. 

        The curve reveals a pronounced peak in JSD when the glitch is injected before 0.04 seconds, during the early phase of the ringdown. This indicates that glitches overlapping with the early, high-amplitude portion of the signal cause the largest bias in parameter estimation. 

        This phenomenon can be attributed to the information content of the ringdown signal, which varies over time. During the early portion of the ringdown, the fundamental mode reaches its highest amplitude and carries most of the signal energy. A glitch overlapping this information-rich phase directly contaminates the strongest part of the waveform, altering its time-frequency structure and distorting the inferred damping rates and mode amplitudes, which leads to the largest biases in the estimates of mass and spin. In the late-time tail, the ringdown amplitude decays exponentially and the instantaneous SNR is low, so the tail contributes comparatively little information to the posterior; glitches injected there produce only modest deviations, with most JSD values remaining below the reference level.
        
        The largest deviations occur for injection times before approximately 0.04 s, where the glitch overlaps the early, high-amplitude portion of the ringdown and interferes with the fundamental mode in the time-frequency domain~\citep{Macas_2022}. 
   
        Consequently, glitches overlapping the early, high-amplitude portion of the ringdown distort the inferred damping rates and mode amplitudes, leading to significant biases in the estimates of mass and spin.
        
        We also assess JSD using the marginal distributions as shown in Fig.~\ref{Time_subplot} below. 
        The marginal JSD trends for mass, spin and phase broadly follow the overall JSD trend, while the amplitude marginal shows larger deviations at late injection times, indicating that the glitch primarily corrupts the amplitude recovery.
        
        \begin{figure}[htbp]
        \centering
        \includegraphics[width=0.48\textwidth]{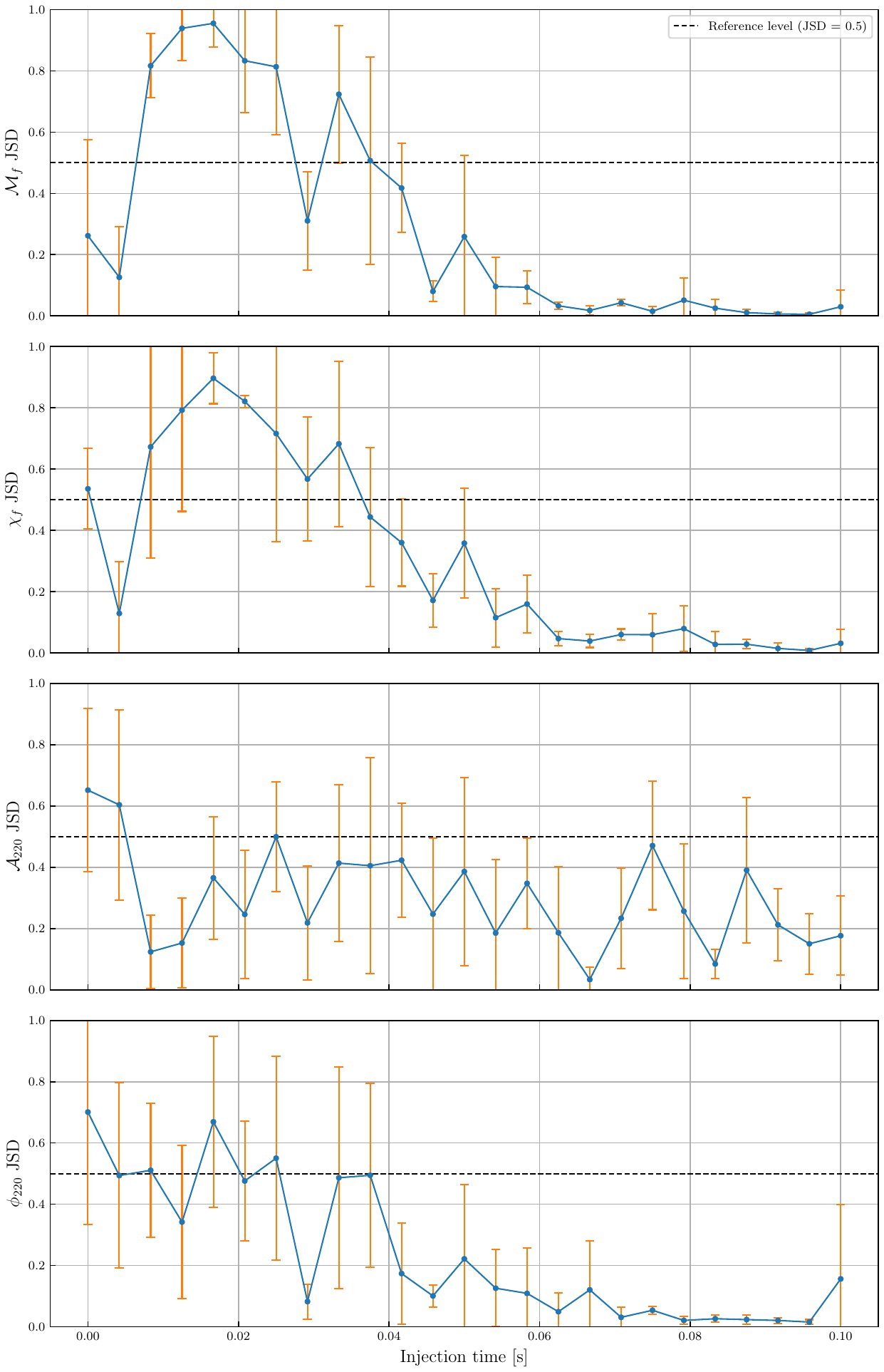}
        \caption{Marginal JSD between the posteriors obtained from glitch-contaminated and clean data at different glitch injection times. The same JSD reference level as in Fig.~\ref{Time_ALL} is shown for reference. }
        \label{Time_subplot}
        \end{figure}
        
    \subsubsection{The SNR of glitches }\label{sec3.2.2}

        \begin{figure*}[htbp]
        \centering
        \includegraphics[width=0.8\textwidth]{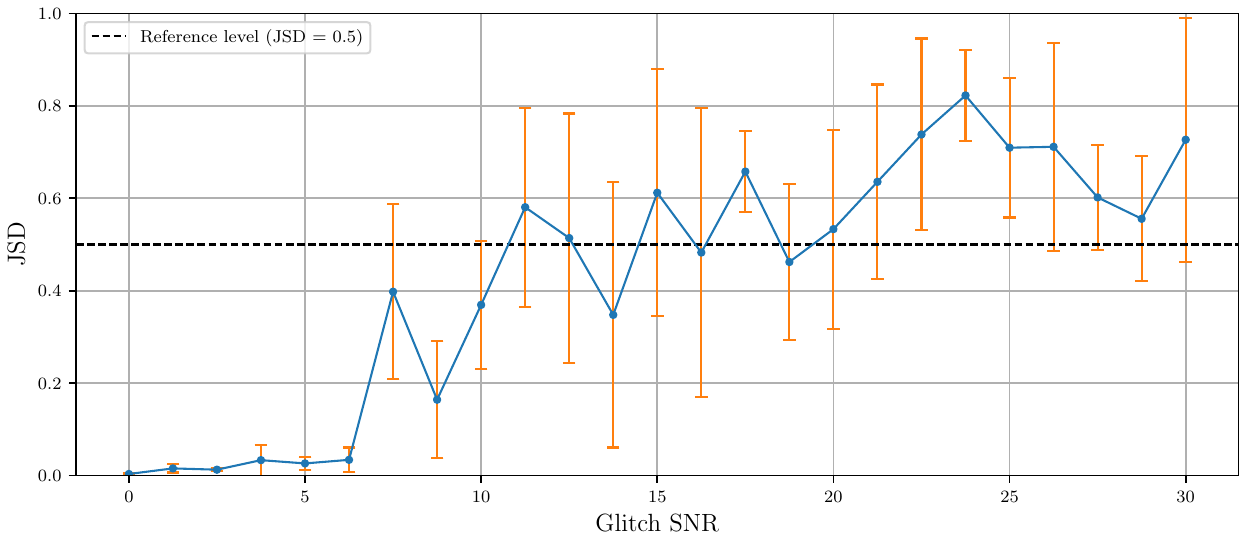}
        \caption{The JSD value between posterior distributions obtained from data contaminated with glitches and those from clean data for different glitches SNR. The data points in the graph are the mean JSD calculated from five independent runs. Error bars represent the SD of these five runs. The same JSD reference level as in Fig.~\ref{Time_ALL} is shown for reference.}
        \label{SNR_all}
        \end{figure*}
        
        Next, we discuss the effect of the SNR of glitches on model posterior inference. In this subsection, the glitch injection time is fixed at the center of the ringdown signal, specifically at $t=0.05$ seconds, and only the glitch SNR is varied.
        Figure~\ref{SNR_all} shows that with increasing glitch SNR, the difference between the posterior distributions of the model and those without noise increases, making the model inference gradually unreliable. 
        
        Under lower SNR conditions, the glitch primarily acts as a background perturbation, and its influence can be partially absorbed by the model's ability to approximate complex distributions. In contrast, under higher SNR conditions, the energy of the glitch approaches or even exceeds that of the main signal, causing the glitch to overlap with the signal modes in the time-frequency domain, thereby confusing mode identification and separation. The glitch has a particularly significant effect on amplitude-dependent and phase-dependent parameters.

        The overall JSD generally increases with glitch SNR, although the trend is noisy and the scatter across the five independent runs is large. 
        
        For glitches injected at this fixed time, the marginal JSDs for mass and spin stay below the reference level of JSD $=0.5$ for all glitch SNRs probed, i.e.\ they remain low in absolute terms even though the mass marginal follows the rising trend of the joint JSD reported above.
        
        A possible reason is that these parameters are mainly constrained by the frequency structure of the ringdown. However, this does not imply general robustness of mass and spin to glitches. 
        
        As shown in Fig. 4, early-time glitches that overlap the high-amplitude part of the ringdown can produce much larger JSDs for mass and spin. Therefore, the stability observed here is specific to the fixed injection time of $t=0.05$ seconds and cannot be generalized to other injection times.
        
        Similarly to the injection time test in Sec.~\ref{sec3.2.1}, we also measure the JSD of marginal distributions as shown in Fig.~\ref{SNR_subplot} below. Overall, all marginal distributions have low JSD values. 
        The JSD trend for mass matches the overall JSD trend, whereas the JSD values for spin and phase parameters show weaker dependence on SNR. However, the marginal distribution of amplitude shows the opposite trend, with lower JSD at high SNR.
        Therefore, we can draw a conclusion that the glitch SNR has a lower impact on the value of JSD than the injection time of glitches.
        
        \begin{figure}[htbp]
        \centering
        \includegraphics[width=0.48\textwidth]{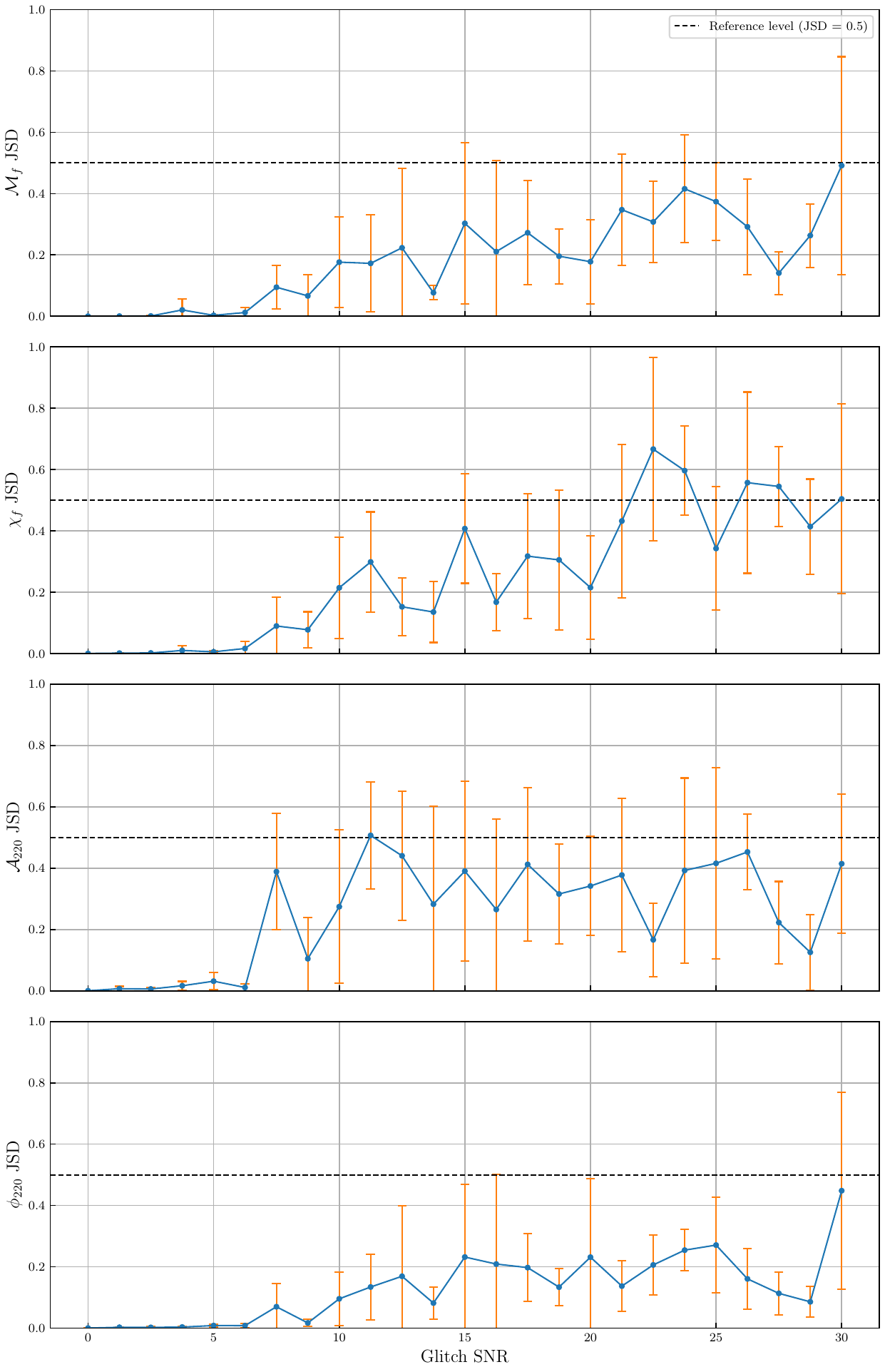}
        \caption{Marginal JSD between the posteriors obtained from glitch-contaminated and clean data at different glitch SNRs. The same JSD reference level as in Fig.~\ref{Time_ALL} is shown for reference. }
        \label{SNR_subplot}
        \end{figure}
        
        
        
        
\section{Conclusion}\label{sec4}

    This study aims to develop and evaluate a likelihood-free method for estimating parameters of GW ringdown signals in the time domain. We focus on its performance under both ideal conditions and non-Gaussian noise interference. By integrating amortized NPE with normalizing flows, we construct an efficient Bayesian inference framework. Using this framework, we systematically evaluate how glitches affect the inference results.  

    The main findings of this study can be summarized in two key areas. Firstly, performance validation on clean data: When sufficiently trained, the amortized NPE model can produce posterior distributions that are statistically consistent with those from classical MCMC methods. Through corner plot comparisons, the KS test, and coverage probability analyses, we have confirmed the model's reliability in terms of parameter recovery capability and statistical calibration. At the same time, the model achieves computational speed-ups of several orders of magnitude, providing a viable solution for processing the high event rates expected from next-generation GW detectors. Secondly, robustness analysis on noise-contaminated data: We have provided systematic quantification of the mechanisms through which glitches affect model performance. Our results demonstrate that the pipeline's performance depends strongly on the conditions, particularly the timing and strength of the glitch.
    
    Our analysis reveals key findings on how glitches affect inference. Firstly, the injection time of a glitch has a significant influence on inference bias. Glitches that overlap with the early, information-rich portion of the signal, where the fundamental mode is strongest, cause the most significant parameter estimation deviations. Secondly, glitch strength correlates positively with inference error. However, this impact remains limited under low SNR conditions. Thirdly, different physical parameters show varying sensitivities to glitches. The impact is parameter dependent, and no single parameter is uniformly the most sensitive across all glitch configurations considered here.
    
    Universal robustness against glitches has not been established in this work. Rather, our findings demonstrate that the pipeline is conditionally reliable: it exhibits only modest deviations for low-strength glitches and for glitches overlapping the late-time, low-information tail, but shows significant bias when glitches overlap the early, high-amplitude portion of the ringdown. Several factors limit the generality of our findings. First, the current work is primarily based on a simplified single-mode signal model ($Kerr_{2,2,0}$) and has not yet been extended to complete waveforms incorporating higher order QNM. Second, the experiments employ simulated glitches whose waveforms may differ from transient noise in real detectors. Third, model training relies heavily on extensive simulated data and is sensitive to the accuracy of both the waveform and noise models. Fourth, and most importantly, we acknowledge that the current study assumes a fixed design PSD for training the density estimator, which is a simplification of real detector conditions where the PSD varies over time.

    Because this study is conditional in scope, our contribution is not a proof that the method works in all cases. Instead, we systematically identify and describe the system's weaknesses, showing when and how the inference pipeline fails when data are contaminated by glitches. This description serves as a useful baseline for developing more robust data-processing pipelines for future GW detectors. We also suggest applying stronger data quality control or model attention mechanisms when the signal carries little information, such as in the ringdown tail.
    
    Future work could extend in the following directions. First, integrating multi-modal signal models to extend the framework to complete ringdown waveforms that include higher order QNM. Second, developing adaptive noise-resistant training strategies. A natural extension is to train the neural estimator on simulated data in which glitches are also injected, treating glitches as a nuisance effect, with the aim of making parameter estimation more robust to their presence. Third, promoting the application of the method in real-time data processing systems to provide efficient tools for online parameter inference in next-generation GW detectors. Finally, to apply amortized inference to real data, a primary strategy is to condition the density estimator on the PSD during training, which allows the network to generalize across different noise realizations. This approach, similar to the methodology used in DINGO~\citep{PhysRevLett.127.241103}, requires a more complex network architecture but preserves the efficiency of inference. While our work provides a systematic diagnostic characterization of vulnerabilities rather than a proof of universal robustness, incorporating PSD conditioning remains a crucial direction for future development to achieve more robust inference.
    
\begin{acknowledgments}

    We thank the developers of the open-source software packages used in this work, including {\tt lampe}, {\tt Pycbc}, {\tt pyRing}, and {\tt cpnest}, which were essential for the implementation and validation of our inference framework. This research has made use of data or software obtained from the Gravitational Wave Open Science Center (gwosc.org), a service of LIGO Laboratory, the LIGO Scientific Collaboration, the Virgo Collaboration, and KAGRA. This work is supported by the National Natural Science Foundation of China (Grants Nos. 12473001, 12575049, and 12533001), the National SKA Program of China (Grants Nos. 2022SKA0110200 and 2022SKA0110203), the China Manned Space Program (Grant No. CMS-CSST-2025-A02), the National 111 Project (Grant No. B16009), and the China Scholarship Council.

\end{acknowledgments}
\appendix
\section{Validation at High-SNR }
\label{app:snr50}

To further validate the inference pipeline at higher signal-to-noise ratio, we repeat the recovery of Sec.~\ref{sec3.1} with the same final mass, spin, and phase but with a larger fundamental-mode amplitude,
\begin{equation}
    \mathcal{A}_{220} = 8.3 \times 10^{-21},
    \label{eq:snr50-amp}
\end{equation}
which yields $\rho \approx 50$ according to the noise-weighted inner product defined in Eq. (4).  The corresponding injection parameters are summarized in Table~\ref{tab:snr50-injection}.  All other settings, including the trained estimators (100-epoch and 200-epoch checkpoints), the pyRing inference settings, and the number of posterior samples, are identical to those described in Sec.~\ref{sec2.3} and Sec.~\ref{sec3.1}.

\begin{table}[htbp]
    \centering
    \caption{Parameters of the high-SNR test injection.}
    \label{tab:snr50-injection}
    \begin{tabular}{lcc}
        \hline
        Parameter & Injection 1 & Injection 2 \\
        \hline
        Target SNR $\rho$      & $\approx 12$ & $\approx 50$ \\
        Total mass $\mathcal{M}_f$ & 67       & 67       \\
        Dimensionless spin $\chi_f$  & 0.67     & 0.67     \\
        Amplitude $\mathcal{A}_{220}$          & $2\times10^{-21}$ & $8.3\times10^{-21}$ \\
        Phase $\phi_{220}$        & $\pi$    & $\pi$    \\
        \hline
    \end{tabular}
\end{table}

\begin{figure*}[htbp]
        	\centering
        	\subfigure[low-epoch]
        	{
        		\begin{minipage}[b]{0.48\linewidth}
        			\centering
        			\includegraphics[scale=0.65]{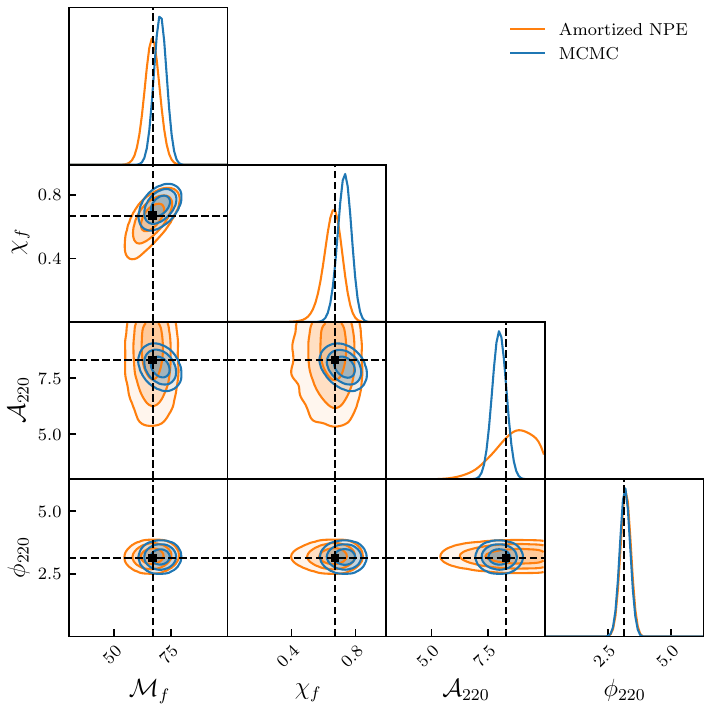}
        		\end{minipage}
        	}
        	\subfigure[high-epoch]
        	{
        		\begin{minipage}[b]{0.48\linewidth}
        			\centering
        			\includegraphics[scale=0.65]{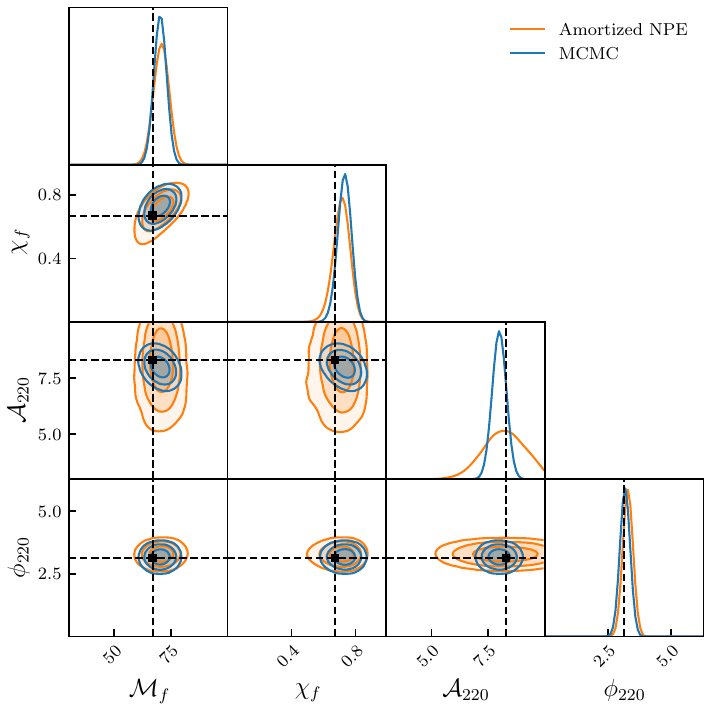}
        	\end{minipage}
        	}
        	\caption{Comparison of parameter estimates and credible intervals from MCMC and the amortized NPE for the $\rho \approx 50$ injection.}
            
        \label{fig:snr50}
        \end{figure*}

The posterior comparison between the amortized NPE and MCMC is shown in Fig.~\ref{fig:snr50}. 

The recovery at $\rho \approx 50$ agrees with the MCMC results in location for $\mathcal{M}_f$, $\chi_f$, and $\phi_{220}$. For $\mathcal{A}_{220}$, the NPE and MCMC posteriors agree in location, but the NPE posterior is broader, indicating that the amplitude is the least tightly constrained parameter; we therefore regard the high-SNR test as a consistency check on parameter location rather than exact agreement for every parameter.

\newpage
\bibliography{paper}

\end{document}